\documentclass[11pt,aps,prd,showpacs,nofootinbib,superscriptaddress,preprint,preprintnumbers]{revtex4}

\usepackage{natbib}
\usepackage{color}
\usepackage{graphicx}
\usepackage{amsmath}
\usepackage[caption=false]{subfig}
\usepackage[colorlinks=true,linkcolor=blue,citecolor=blue,urlcolor=black]{hyperref}

\newcommand{\be}{\begin{equation}}
\newcommand{\ee}{\end{equation}}
\newcommand{\bea}{\begin{eqnarray}}
\newcommand{\eea}{\end{eqnarray}}
\newcommand{\ppbbh}{$pp\to b \bar b H+X\,$}

\newcommand{\GeV}{\,\rm{GeV}}
\newcommand{\MeV}{\,\rm{MeV}}
\newcommand{\fb}{\,\rm{fb}}
\newcommand{\nn}{\nonumber}

\usepackage{ulem,fancyvrb}

\usepackage{xcolor}
\usepackage{lipsum}
\usepackage{multirow}

\begin{document}
\title{The NLO electroweak effects on the Higgs production in association with bottom quark pair  at the LHC}
\author{Yu Zhang} \email{dayu@nju.edu.cn}
\affiliation{School of Physics, Nanjing University, 
 Nanjing, Jiangsu, 210093, China}
\affiliation{CAS Center for Excellence in Particle Physics, Beijing 100049, China}

\begin{abstract}
The dominant contribution to the Higgs production in association with bottom quark pair  at the LHC
is gluon-gluon fusion parton subprocess. We present a complete calculation of 
the next-to-leading order (NLO) electroweak (EW) corrections to this channel. The other small
contributions with quarks in the initial state are calculated at tree level.
We find that the NLO EW corrections can suppress the leading order (LO) contributions significantly.
\end{abstract}

\maketitle


\section{Introduction \label{sec:intro}}
\par
The discovery of the long waited Higgs boson at the LHC by ATLAS\cite{Aad:2012tfa} and CMS\cite{Chatrchyan:2012ufa} in July 2012 
is a milestone in the particle physics. After this achievement, more precise
examination of this new boson's properties becomes one of the most 
important endeavors for the LHC and future colliders.
The recent analyses show that its couplings are compatible with those predicted
by the Standard Model(SM)\cite{atlas:2013sla,Chatrchyan:2013lba}.
However, the interpretation of Beyond the Standard Model(BSM) scenarios is still an open issue and more precise predictions 
for this particle are urgently required. 

\par
The production of a Higgs boson in association with bottom quarks at hadron colliders has been extensively studied in the literature.
Depending  on  the  choice  of  the  flavour-scheme  in  the  partonic  description  of  the  initial
state  and  on  the  identified  final  state,  one  can  consider  a  number  of  different  partonic
sub-processes for associated bottom-Higgs production: while the choice of the 4 versus 5 flavour scheme is
mainly theoretically motivated, resulting in a reordering of the perturbative expansion \cite{Campbell:2004pu},
the requirement of a minimum number of tagged $b$ in the final state is physically relevant in
the signal extraction. There are mainly three  different types of production processes: i) the  inclusive  one,  where  no
bottom quark jet is tagged, dominated by the bottom quark fusion process $b\bar b\to H$, ii) the semi-inclusive one,
where only one bottom quark is tagged, dominated by process $bg\to bH$, iii) the exclusive one where both bottom jets are tagged,
almost entirely dominated by $gg\to b\bar b H$, with only a small contribution from $q\bar q\to b\bar b H$.

\par
The analysis of the relative weights of the above three different types of production processes are present in \cite{Dicus:1998hs,
Balazs:1998sb},
where also the $b\bar b\to H$ process is computed at next-to-leading order (NLO) in QCD. Besides, the $b\bar b\to H$ process
has been calculated at next-next-to-leading order (NNLO) in QCD \cite{Harlander:2003ai} 
while the electroweak NLO corrections have been presented in \cite{Dittmaier:2006cz}.
For the  associated  semi-inclusive  production  process $bg\to bH$, NLO QCD corrections can be obtained from 
\cite{Campbell:2002zm,Dawson:2004sh,Dawson:2007ur},
and purely-weak and EW corrections have been presented in \cite{Dawson:2010yz,Beccaria:2010fg} respectively. 
Finally, for the  exclusive  process,  where  two  bottom  jets  are  tagged  in  the  final  state,  
the cross section is known through NLO QCD in the SM in the four-flavor scheme (4FS)
\cite{Dittmaier:2003ej, Dawson:2003kb, Dawson:2005vi} and matched to parton showers in \cite{Wiesemann:2014ioa}.  

\par
This paper is strongly motivated by the possible relevance of the associated bottom-Higgs
production in the experimental examination of the bottom-quark Yukawa couplings at the LHC.
The purpose of this paper is to provide and study the EW corrections to the fully exclusive process
$pp\to b\bar b h$ in the 4FS, where the final state includes two high transverse momentum bottom quarks, for the first time.
The rest of this paper is organized as follows:
in section \ref{sec:cal}, we describe the structure of EW NLO calculation for the $b \bar{b} H$ production at the LHC.
The numerical results are
presented and discussed in section \ref{sec:numerical}. Finally a summary is given.                                                                                 
\section{Structure of the calculation}
\label{sec:cal}

\subsection{Leading order consideration}
\label{sec:LO}

\par
\begin{figure*}
\begin{center}
\includegraphics[scale=0.6]{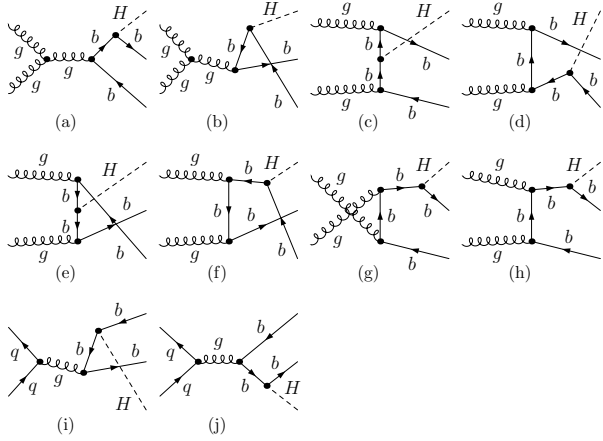} 
\caption{The Feynman diagrams at tree level. 
     } 
\label{fig:lo}
\end{center}
\end{figure*}

For the process \ppbbh, at tree level the main partonic subprocesses are $gg\to b\bar b H$ and
$q\bar q\to b\bar b H$ ($q$ stands for light quark). The corresponding
Feynman diagrams are displayed in the first two lines (a-h) and the last line (i-j) of Fig.\ref{fig:lo},
respectively.
The dominant contributions arise from $gg$ of 
order ${\cal O} (\alpha_s^2\alpha)$, and the $q\bar q$ contributions of the same order are 
much smaller since they involve only the $s$ channel diagrams, which are 
obviously suppressed at high energy, and the corresponding parton density is much smaller
than that of gluon at the LHC. It is noticed that there are also contributions of the order 
${\cal O} (\alpha^3)$ from the tree-level EW Feynman diagrams of $q\bar q$ annihilations,
are neglected in our calculation since they are extremely small.

\par
At LO (${\cal O} (\alpha_s^2\alpha)$), the total cross section for \ppbbh can be written as  
\begin{align}
 \sigma_{LO}^{pp} & =  \sum_{ij}\frac{1}{1+\delta_{ij}}\int dx_1 dx_2
 \left[  {\cal F}_i^p(x_1,\mu_F){\cal
   F}_j^p(x_2,\mu_F)\hat{\sigma}_{LO}^{ij}(x_1,x_2,\mu_R)+(x_1\leftrightarrow x_2)
 \right],
 \end{align}
where $(i,j)$ denotes $(g,g)\,(q,\bar{q})$, $\hat{\sigma}_{LO}^{ij}$
the corresponding partonic cross section at LO and ${\cal F}^p_{j}(x,\mu_F)$
represents the distribution function at the scale $\mu_F$ of parton $i$(i.e., 
quark or gluon ) at momentum fraction of $x$. 

\subsection{ NLO EW contribution to $gg \to b \bar{b}H$ production }
\label{sec:EW}

\par
In the following, we discuss the NLO EW contributions to the $gg \to b \bar{b}H$ production.
These contributions are of order ${\cal O}(\alpha_s^2 \alpha^2)$.
The contributions at the same order from the suppressed $q\bar q$ annihilations 
are much smaller and will be neglected.

\par
The NLO EW correction includes both the virtual and real photon contribution. 
The virtual corrections are induced by the self-energies, triangle (3-point), 
box (4-point), and pentagon (5-point) diagrams, 
and contain ultraviolet(UV) divergences and infrared (IR) soft 
divergences.
First, we discuss the issue of renormalization which guarantees the result
ultraviolet safe. 
It should be noted that the box and pentagon diagrams are both UV finite.
To cancel the UV divergences, the renormalization of the bottom quark mass, the
Higgs mass, the electric coupling, the bottom Yukawa coupling, and the external wave 
functions are required. The relevant wave functions and masses are renormalized 
by taking the on-mass-shell renormalization scheme. The renormalization 
constants can be found in Ref.\cite{Denner:1991kt}.
To specify the fine structure constant, we adopt the $G_\mu$ scheme with 
$\alpha_{G_\mu} = \frac{\sqrt{2}M_W^2 G_{\mu}}{\pi}(1-\frac{M_W^2}{M_Z^2}) $
as an input parameter. With this choice, the EW corrections are independent of 
logarithms of the light fermion masses, and the calculation is consistently done
by modifying
the renormalization constant according to 
\begin{align}
  \delta Z_e^{G_\mu} = -\frac{1}{2} \delta Z_{AA} - \frac{s_W}{2c_W} \delta Z_{ZA} -\frac{1}{2} \Delta r^{(1)} \, ,
\end{align}
where the explicit expressions of $\Delta r^{(1)}$ are detailed in Ref.\cite{Denner:1991kt}.
Concerning to bottom quark, the pole mass enters the kinematic variables of the matrix element and the 
phase space, and a running bottom mass is usually used in the improved Higgs Yukawa coupling. 
For NLO QCD calculations, the two masses can be treated as different variables. 
However, as the bottom mass is of EW origin, this treatment is not feasible for NLO EW analysis.
Such treatment would violate Ward identities involving $m_b$\cite{Baro:2008bg}, and the cancellation
of UV poles will be incomplete. Consequently, one has to implement a common value, either 
the pole mass or the running mass, for the 
bottom quark mass that enters the kinematic variables of the matrix element, 
the phase space and the variable of Higgs-bottom Yukawa couplings\cite{Nhung:2012er}. In our calculation, 
the cancellation of UV poles are checked by using the bottom pole mass as a common value.

\par
\begin{figure*}
\begin{center}
\includegraphics[scale=0.7]{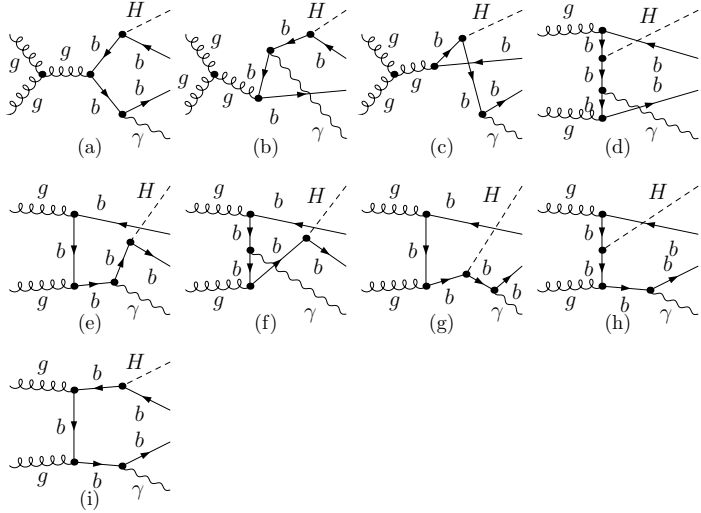}
\caption{The representative Feynman diagrams for the real photon emission
process $gg \to  b \bar{b} H+ \gamma$. }
\label{fig:rA}
\end{center}
\end{figure*}

Eliminating the UV divergences, the virtual corrections still involve soft IR divergences.
In our consideration, to obtain an IR safe result, we need to take
into account the real photon emission contributions arise from the subprocess: 
\begin{equation}
g+g \to b+\bar{b} + H+ \gamma,
\label{eq:gghbba}
\end{equation}
of which the representative Feynman diagrams are shown in Fig.\ref{fig:rA}.
The photon induced IR singularities originating from the virtual
corrections can be extracted and cancelled
exactly with those in the real photon emission corrections. This cancellation 
can be realized either by the two cutoff phase space slicing
method \cite{Harris:2001sx} or the dipole subtraction method \cite{Catani:1996jh, Catani:1996vz,Catani:2002hc}.
In this work, we adopt the two cutoff phase space slicing method with $\delta_s=1\times10^{-5}$ and verified the
result is consistence with the result by using dipole subtraction. 

\par
\begin{figure*}
\begin{center}
\includegraphics[scale=0.7]{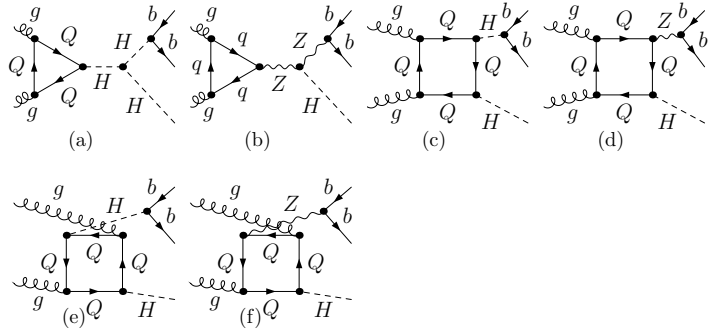}
\caption{The representative one-loop diagrams with possible resonance for the
process $gg \to  b \bar{b} H$. $Q=t,b$. $q=u,d,c,s,t,b$.}
\label{fig:res}
\end{center}
\end{figure*}

Since the invariant mass of the bottom pair can equal to Higgs or Z
boson mass, the intermediate Higgs or Z boson
can be on-shell, which is illustrated in Fig.\ref{fig:res}. The interference between the corresponding diagrams and the 
LO diagrams would contain a resonant propagator $\frac{1}{M_{b\bar b}^2-M_{(H/Z)}^2}$, which leads to singularities
in the vicinity of $M_{b\bar b}^2 \sim M_{(H/Z)}^2$. We regulate the singularities by making the replacement of
$\frac{1}{M_{b\bar b}^2-M_{(H/Z)}^2}\rightarrow
\frac{1}{M_{b\bar b}^2-M_{(H/Z)}^2+iM_{(H/Z)}\Gamma_{(H/Z)}}$. The corresponding contribution
is found to be negligible in the total NLO EW cross section, but it is important for several differential kinematic distributions.

We apply the FeynArts-3.7 package \cite{Hahn:2000kx} to generate the Feynman diagrams automatically
and the corresponding amplitudes are algebraically simplified by the FormCalc-7.2 program \cite{Hahn:1998yk}.
In the calculation of one-loop Feynman amplitudes, the {\sc LoopTools-2.8}
package \cite{Hahn:1998yk} is adopted for the numerical calculations of the scalar and tensor integrals, in which the $n$-point ($n\le 4$)
tensor integrals are reduced to scalar integrals recursively by Passarino-Veltman algorithm and the 5-point integrals are
decomposed into 4-point integrals by the method of Denner and Dittmaier \cite{Denner:2002ii}.

\section{Numerical studies}
\label{sec:numerical}

\subsection{Input parameters}
\label{sec:parameters}

\par
For the numerical analysis we take the following input parameters \cite{Patrignani:2016xqp},
\begin{alignat}{2}
  G_{\mu} & =  1.1663787 \times 10^{-5} \GeV^{-2},  &\hspace{10pt}     M_W & =  80.385\GeV,  \nn \\
  M_Z & =  91.1876\GeV,                       &               \Gamma_{Z} & =  2.4952 \MeV,  \nn \\
      M_t & =  173.1\,\GeV,                       &                  M_H & =  125\GeV .
  \label{eq:SMpar}
\end{alignat}
  \vspace{2ex}
We adopt the MSTWlo2008 PDFs \cite{Martin:2009iq} with 4-flavor light quarks in the numerical analysis
for NLO as well as LO predictions (since we are chiefly interested in assessing effects of matrix-element
origin). If not
otherwise specified, the
renormalization scale and the factorization scale are set to be equal, that is,
$\mu_R\,=\,\mu_F\,=\,(M_H+2\times m_b)/4$. 
As explained in section \ref{sec:EW}, the bottom quark mass is set as pole mass with 4.78 GeV.
The decay width of Higgs boson is assumed to be $\Gamma_{H}\;=\;6\;\text{MeV}$\cite{Barger:2012hv}.

\par
Both the final state bottom and anti-bottom quarks is tagged by the following kinematic constraints:
\begin{align}\label{eq:cut}
 p_T^{b/\bar b}~>~30 ~{\rm GeV}, ~~    |y^{b/\bar b}|~ <~ 2.5, ~~  
 R_{b\bar b}~ >~ 0.4,
\end{align}   
where
$p_T^{b/\bar b}$ and $y^{b/\bar b}$ are the transverse momentum and
rapidity of the bottom and anti-bottom quarks, respectively, and 
$R_{b\bar b}$ is the separation in the plane of azimuthal angle and
rapidity between two $b$ quarks.

\subsection{Total cross sections}
\label{sec:total}

\par
From the description in above section, we can obtain the total EW NLO corrected 
cross section for $pp \to  b \bar{b} H$ process as
\begin{align}
  \sigma_{NLO}^{pp}& = \sigma_{LO}^{pp} + \Delta
  \sigma_{EW}^{gg} ,
  \label{eq:ew}
\end{align}
where $\Delta \sigma_{EW}^{gg}$ is the summation 
of the virtual and real photon corrections for the subprocess 
$gg \to  b \bar{b} H$.

\par
We show the total LO and EW NLO cross section at the 13 TeV
LHC in Table \ref{tb:total} for some typical values of the factorization/renormalization 
scale, where $\mu_R\,=\,\mu_F\,=\mu$ and $\mu_0\,=\,(M_H+2\times m_b)/4$. 
The corresponding relative corrections $\delta$ in the last column are defined as
$\delta\,=\,(\sigma_{NLO}^{pp}-\sigma_{LO}^{pp})/\sigma_{LO}^{pp}$.
It's worth noting that, all the 
contributions of $q\bar q$ annihilations of the giving values of the factorization/renormalization 
scale are less $2\%$ to the total cross section at LO.
This is the very motivation that we only include the
NLO EW correction of the $gg$ subprocess.
We can see that the EW NLO relative corrections don't vary with the factorization/renormalization
scale, and the LO cross sections are suppressed by EW NLO corrections by about $4\%$ at the 13 TeV LHC.
To investigating the EW NLO contributions in variations of the bottom pole mass, we present the numerical 
results for typical values of bottom pole mass with  upper value of $4.84$ GeV, center value of $4.78$ GeV 
and lower value $4.72$ GeV \cite{Patrignani:2016xqp} in Table \ref{tb:mb}.
We find that the EW NLO relative corrections remain the same with different bottom pole masses. 

The integrated luminosity $L$ of LHC will reach 300 fb$^{-1}$  in its first 13-15 years of operation,
then LHC will be substituted by the High Luminosity LHC (HL-LHC) with 3000 fb$^{-1}$ \cite{Apollinari:2017cqg}.
The $b\bar b H$ associate production (with $H\to b\bar b$) at LHC has been proposed and systematically studied in \cite{Balazs:1998nt}.
The EW NLO corrections to the event number of $pp\to  b\bar b H\to b\bar b b\bar b$ at the LHC/HL-LHC 
can be simply estimated by
\be
\Delta N = (\sigma_{NLO}^{pp}-\sigma_{LO}^{pp})\times  Br(H\to b\bar b)\times L \times \epsilon_b^4,
\ee
where the branch ratio $Br(H\to b\bar b)=58\%$ \cite{Barger:2012hv} and b-tagging efficiency $\epsilon_b=77\%$ \cite{b-tagging}.
We can find that the EW NLO corrections reduce the event number by 48 at the 13 TeV LHC with 300 fb$^{-1}$ for $\mu_R\,=\,\mu_F\,=\mu_0$,
which  can not be negligible roughly.
\begin{table}
\begin{center}
\begin{tabular}{ c|c|c|c|c|c}
\hline \hline  
$\mu$   & $\sigma_{LO}^{gg}[\fb]$ & $\sigma_{LO}^{qq}[\fb]$
&$\sigma_{LO}^{pp}[\fb]$ &$\sigma_{NLO}^{pp}[\fb]$  & $\delta[\%]$  
\\ \hline
  $\mu_0/4$  & 41.96(4) & 0.6064(1)& 42.57(4) &  41.06(4)  &  -3.6   \\
  $\mu_0/2$  & 29.82(3) & 0.4422(1)& 30.26(3) &  29.18(3)  &  -3.6  \\
  $\mu_0$  & 21.62(2) & 0.3354(1)& 21.96(2) &  21.18(2)  &  -3.6   \\
  $2\mu_0$  & 16.06(1) & 0.26202(8)& 16.32(1) &  15.74(1)  &  -3.6   \\
  $4\mu_0$  & 12.21(1) & 0.20949(6)& 12.42(1) &  11.97(1)  &  -3.6   \\
\hline \hline
\end{tabular}
\caption{The LO, NLO EW corrected integrated cross sections ($\sigma_{LO}$,
$\sigma_{NLO}$ ) and the corresponding
$\delta$ at the 13 TeV LHC
for some typical values of the factorization/renormalization
scale, where $\mu_R\,=\,\mu_F\,=\mu$ and $\mu_0\,=\,(M_H+2\times m_b)/4$.
}
\label{tb:total}
\end{center}
\end{table}

\begin{table}
\begin{center}
\begin{tabular}{ c|c|c|c|c|c}
\hline \hline  
$m_b$ (GeV)   & $\sigma_{LO}^{gg}[\fb]$ & $\sigma_{LO}^{qq}[\fb]$
&$\sigma_{LO}^{pp}[\fb]$ &$\sigma_{NLO}^{pp}[\fb]$  & $\delta[\%]$  
\\ \hline
  $4.84$  & 22.14(2) & 0.3438(1)& 22.49(2) &  21.69(2)  &  -3.6   \\
  $4.78$  & 21.62(2) & 0.3354(1)& 21.96(2) &  21.18(2)  &  -3.6   \\
  $4.72$  & 21.11(2) & 0.3271(1)& 21.44(2) &  20.67(2)  &  -3.6   \\
\hline \hline
\end{tabular}
\caption{The LO, NLO EW corrected integrated cross sections ($\sigma_{LO}$,
$\sigma_{NLO}$ ) and the corresponding
$\delta$ at the 13 TeV LHC with different bottom pole mass.}
\label{tb:mb}
\end{center}
\end{table}

\subsection{Differential cross sections}
\label{sec:distribution}

\par
In the following, we turn to the differential distributions of various
kinematic variables at the 13 TeV LHC.
The relative NLO EW corrections to the differential cross
section $d\sigma/dx$ are defined as $\delta(x)=\left(\frac{d\sigma_{NLO}}{dx}-\frac{\sigma_{LO}}{dx}\right)/\frac{\sigma_{LO}}{dx}$,
where $x$ stands for kinematic observable, i.e., the transverse momentum of the bottom quark ($p_T^b$)
and the Higgs boson ($p_T^H$), the invariant mass of $b\bar b$ pair ($M_{b\bar b}$) 
and the separation between two bottom quarks ($R_{b\bar b}$) in this paper.

\begin{figure*}
\begin{center}
\includegraphics[angle=0,width=3.2in,height=2.4in]{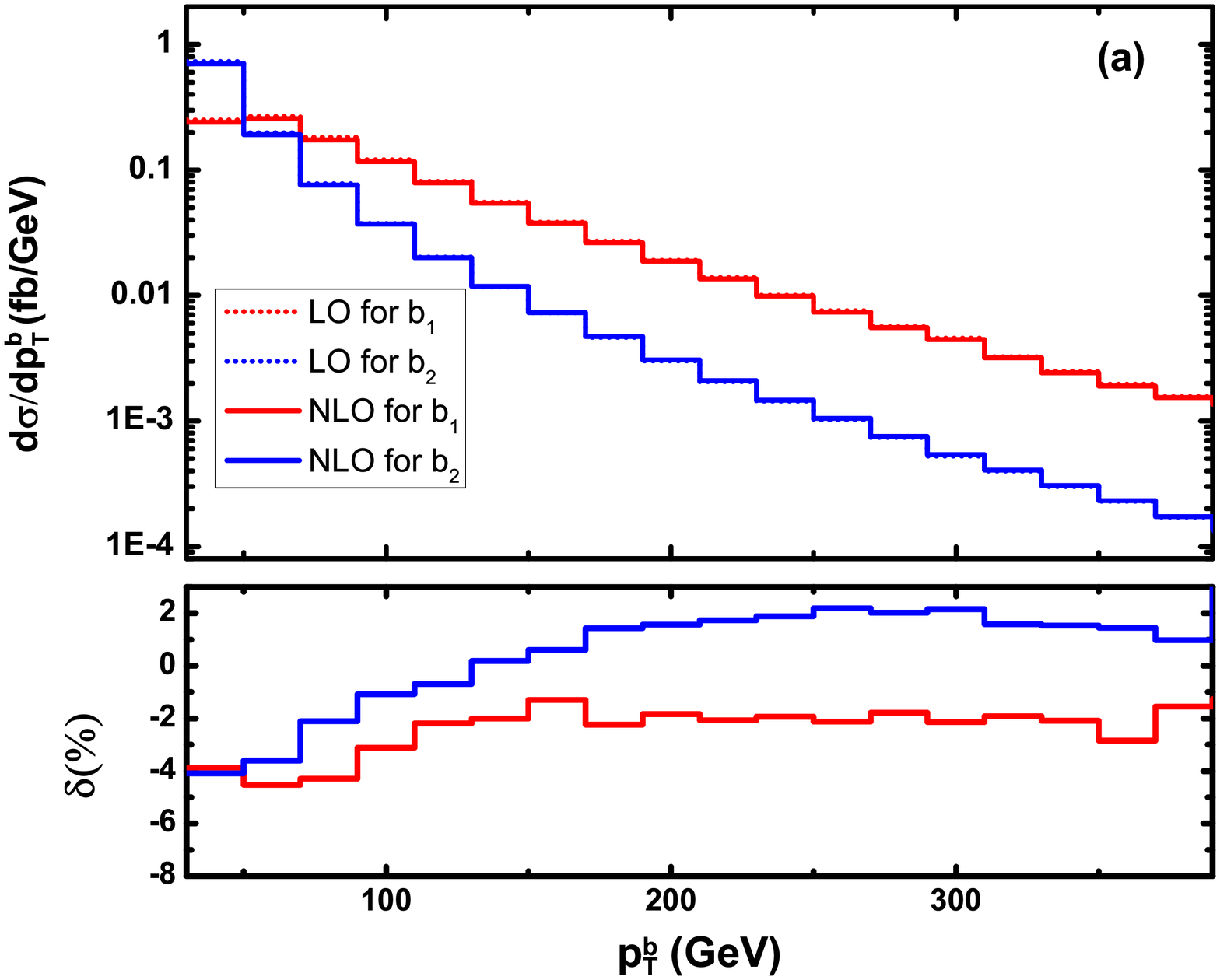}
\includegraphics[angle=0,width=3.2in,height=2.4in]{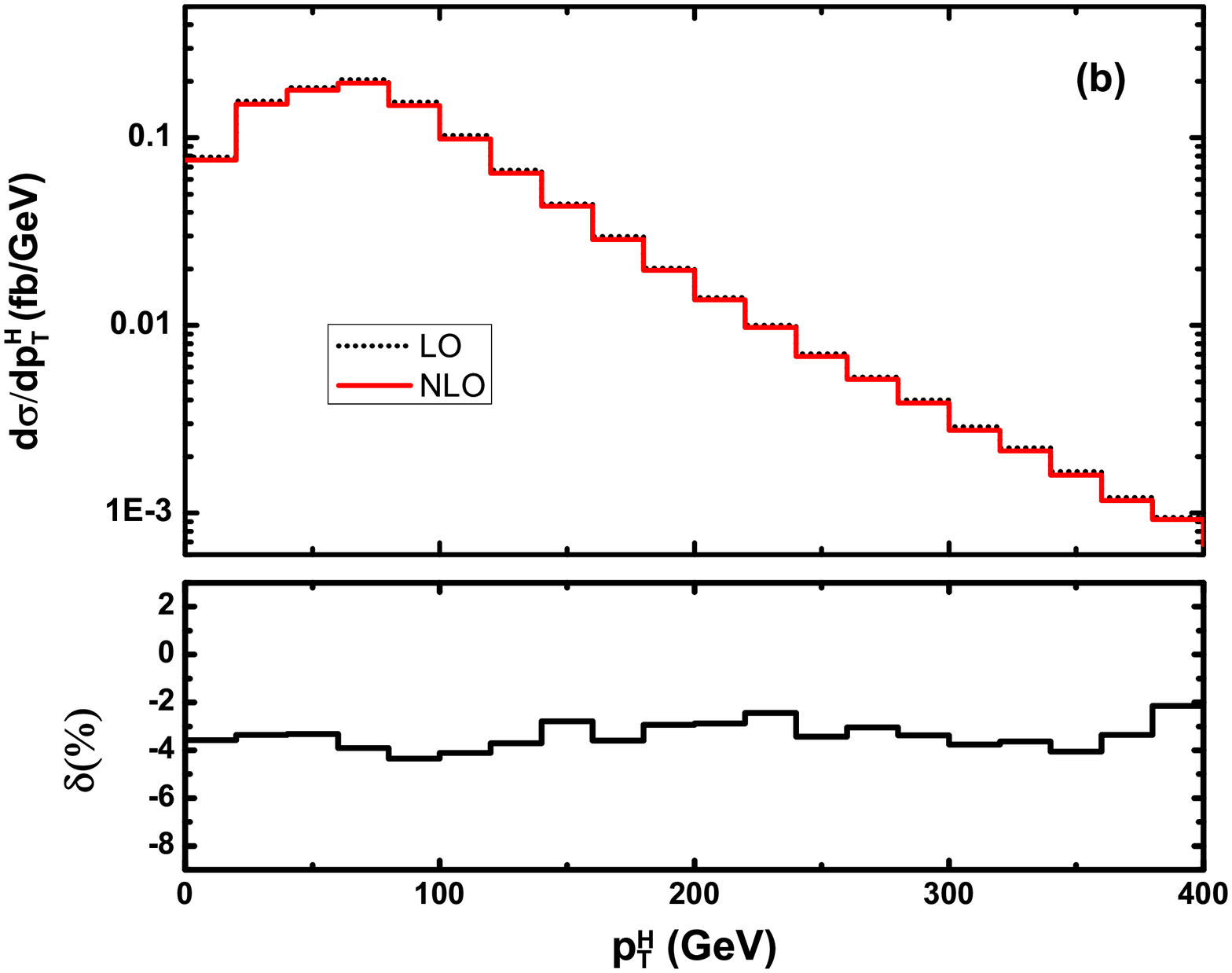}
\caption{The LO, NLO EW corrected distributions and the relative NLO EW
corrections of \ppbbh process at the 13 TeV
LHC for $p_T^b$ (a) and $p_T^H$ (b).  
}
\label{fig:pt}
\end{center}
\end{figure*}

\begin{figure*}
\begin{center}
\includegraphics[angle=0,width=3.2in,height=2.4in]{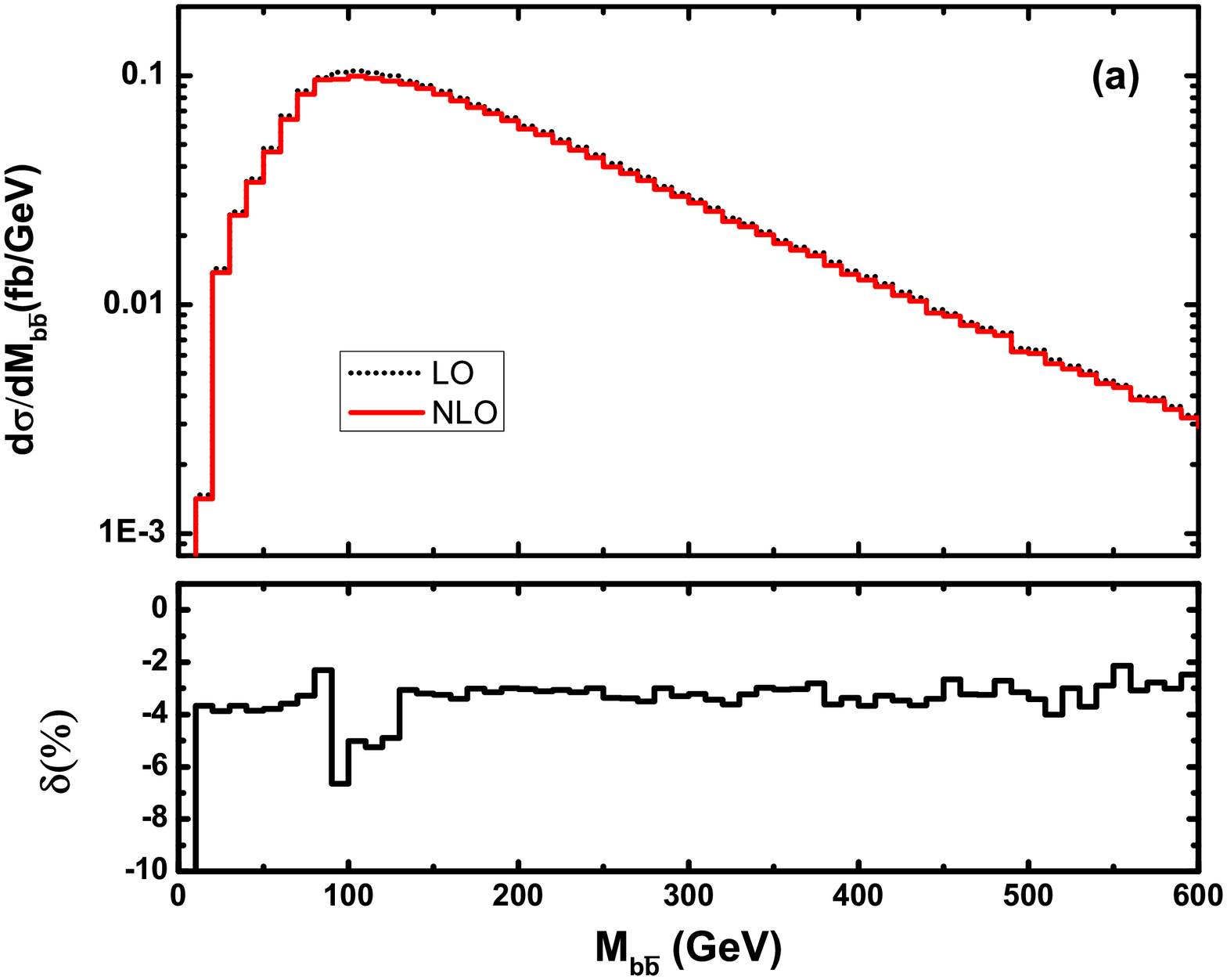}
\includegraphics[angle=0,width=3.2in,height=2.4in]{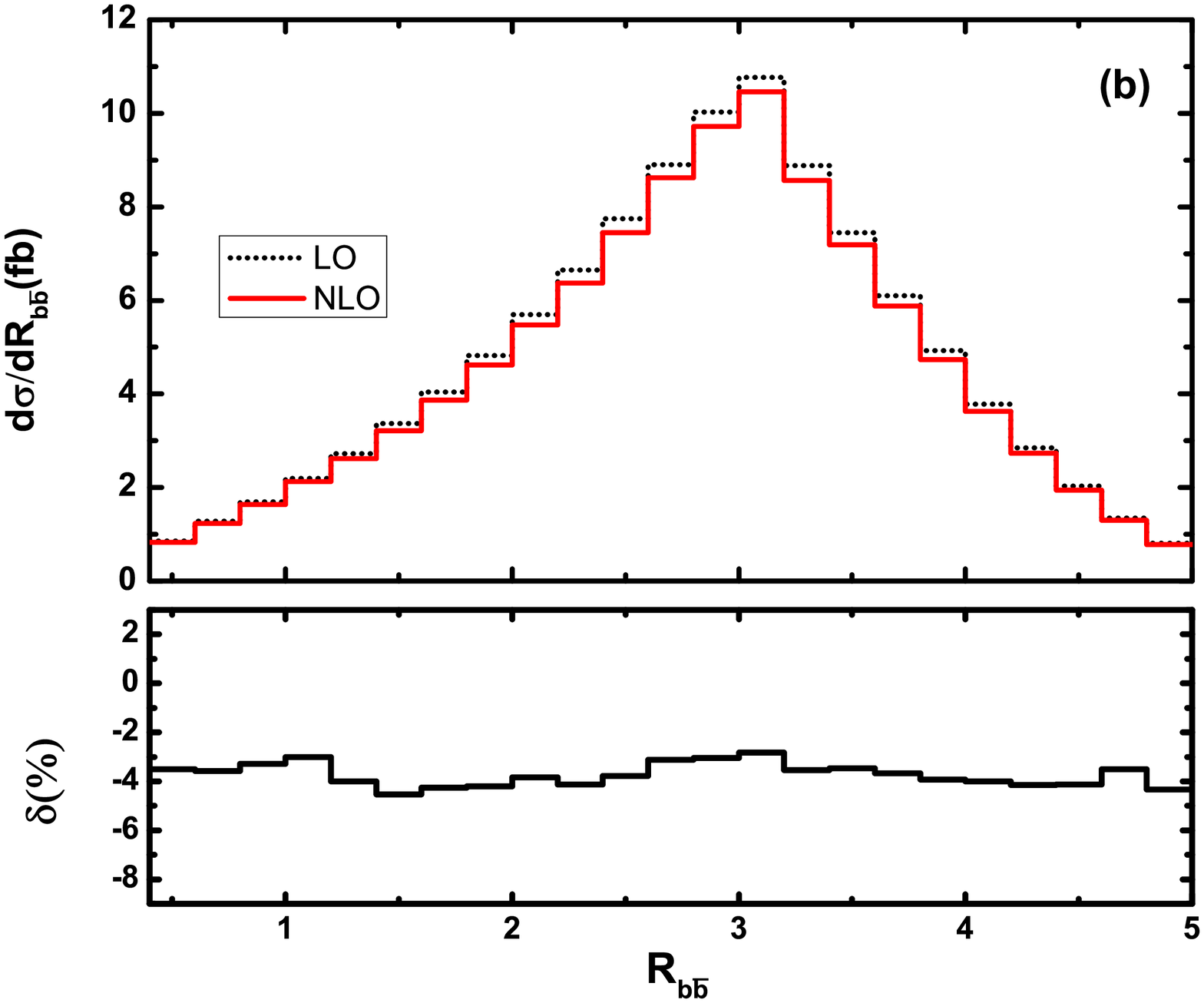}
\caption{The LO, NLO EW corrected distributions and the relative NLO EW
corrections of \ppbbh process at the 13 TeV
LHC for $M_{b\bar b}$ (a) and $R_{b\bar b}$ (b).        
}
\label{fig:mr}
\end{center}
\end{figure*}

In Fig.\ref{fig:pt}(a) and (b), we depict the LO and NLO EW corrected distributions for the
transverse momentum of the final state bottom or anti-bottom  quark with highest (leading)  and second highest (sub-leading) $p_T$
and the Higgs boson. The leading (sub-leading)  b-jet is labelled by $b_1$($b_2$) and colored red (blue).
The Fig.\ref{fig:mr}(a) and (b) present the LO and NLO EW corrected distributions of
$M_{b\bar b}$ and $R_{b\bar b}$ separately. The corresponding relative NLO EW corrections are also shown.
We can see that all the LO distributions of the considered observables (except $p_T^{b_2}$) 
are suppressed by the EW corrections in the plotted region. 
It can be also seen that the EW corrections don't vary the shape of the LO distributions.
For the transverse momentum of the leading and sub-leading b-jets, the LO and EW corrected distributions 
both always decrease with the increment of $p_T^b$ and the relative EW corrections
can be about $-4.5\%$ ($-4.1\%$) in the low $p_T^{b_1}$ ($p_T^{b_2}$) region.
For the transverse momentum of the Higgs boson, all the LO and EW corrected distributions 
reach maximum when $p_T^H\sim60$ GeV and the corresponding relative EW corrections can mount to 
about $-3.9\%$.
From the curve of relative EW corrections of the invariant mass of the bottom quark pair, we can find an obvious
oscillation.
This oscillation shows the singular pole 
structure we mentioned in \ref{sec:EW}: as the invariant mass of bottom quark
pair approaches the Higgs boson mass or the $Z$ boson mass, there exist resonant 
propagators in the virtual diagrams, which lead the oscillation in the
invariant mass distributions.
All the  LO and EW corrected distributions of $R_{b\bar b}$ have peak when $R_{b\bar b}\sim3$,
and the values of relative EW corrections do not vary too much and almost lie in the range of 
$[-3\%\sim-5\%]$ in the plotted region.

\section{Summary}

\par
In this work we have analysed the EW effects for the Higgs boson production 
associated with two bottom quarks at the LHC.
From the numerical analysis, we noticed that at LO ( i.e., $\alpha_s^2
\alpha$) the $gg$ channel is the dominant contribution compared with the other
subprocesses initiated by quark pair. Based on the fact of this 
observation, we only include the EW NLO corrections for the $gg$ channel.
We present the calculation with  some typical values of the factorization/renormalization scale
and bottom pole mass, and find that the EW NLO relative corrections are all $-3.6\%$.
We also investigate the various kinematic distributions of the final state particles, the bottom and 
anti-bottom quarks and the Higgs boson and find that  NLO EW corrections are always negative except in some regions of $p_T^{b_2}$. 

\section*{Acknowledgements}
This work was supported by the China Postdoctoral Science Foundation (Grant No.2017M611771).

\end{document}